\documentclass[aps,pre,twocolumn,a4paper,groupedaddress,superscriptaddress,showpacs,preprintnumbers,amsmath,amssymb,floatfix]{revtex4-1}

\usepackage{graphicx}
\usepackage{stackrel}

\newlength{\onefig}
\newlength{\twofig}
\usepackage[usenames,dvipsnames]{color}

\newcommand{\lps}{\text{LPS}}
\newcommand{\etal}{\textit{et al.}}

\begin{document}

\title{Locally preferred structures and many-body static correlations in viscous liquids}

\author{Daniele Coslovich}
\email{Email: daniele.coslovich@univ-montp2.fr}
\affiliation{Laboratoire Charles Coulomb UMR 5221, Universit\'e Montpellier II
  et CNRS, Montpellier, France}
\affiliation{Institut f\"ur Theoretische Physik and CMS, Technische Universit\"at Wien, Vienna, Austria} 

\date{\today}

\begin{abstract}
  We investigate the influence of static correlations beyond the pair
  level on the dynamics of selected model glass-formers. We compare
  the pair structure, angular distribution functions, and statistics of
  Voronoi polyhedra of two well-known Lennard-Jones mixtures as well
  as of the corresponding Weeks-Chandler-Andersen variants, in which
  the attractive part of the potential is truncated. By means of the
  Voronoi construction we identify the atomic arrangements
  corresponding to the locally preferred structures of the models. We
  find that the growth of domains formed by interconnected locally
  preferred structures signals the onset of the slow dynamics regime
  and allows to rationalize the different dynamic behaviors of the
  models. At low temperature, the spatial extension of the
  structurally correlated domains, evaluated at fixed relaxation time,
  increases with the fragility of the models and is systematically
  reduced by truncating the attractions. In view of these results,
  proper inclusion of many-body static correlations in theories of the
  glass transition appears crucial for the description of the dynamics
  of fragile glass-formers.

\end{abstract}


\maketitle

\section{Introduction}

The question of whether the structure of a glass differs from that of
the corresponding liquid is often rhetorically posed within the glass
community. In fact, very small differences are observed in the static
structure factor of a viscous liquid as it approaches the glass
transition temperature $T_g$. By contrast, the viscosity and
structural relaxation times increase by several orders of magnitude
upon supercooling and the motion of the molecules in the liquid
becomes increasingly cooperative and spatially heterogeneous. 

At first glance, the small structural changes discernible at the level
of pair correlations appear insufficient to explain the dramatic
slowing down of the liquid and the non-trivial spatial correlations of
the dynamics. A preliminary indication that this may not
necessarily be the case already comes from the Mode Coupling theory
(MCT) of the glass transition~\cite{gotze__1999}. The MCT predictions
for the dynamic correlation functions are based uniquely on structural
information, almost invariably the static structure factors of the
liquid. Numerical solutions of the MCT equations for model liquids
show that small variations of the pair correlation functions, which
develop upon lowering the temperature, can produce significant effects
on the dynamics and eventually lead to an ergodic--non-ergodic
transition at some critical temperature
$T_\text{MCT}$~\cite{nauroth_quantitative_1997}. The generic
predictions of the theory account rather well for experimental and
numerical findings in weakly supercooled
liquids~\cite{gotze__1999,das_mode-coupling_2004} (i.e., for $T\agt
T_\text{MCT}$). The break-down of MCT at lower temperatures ($T_{g}< T
\alt T_\text{MCT}$), where the actual system remains effectively
ergodic, most likely reflects the mean-field character of the
theory~\cite{brumer_mean-field_2004,andreanov_biroli_bouchaud_2009,ikeda_mode-coupling_2010}
and its inability to describe activated transitions between metastable
glassy states~\cite{lubchenko_theory_2007}. Another delicate aspect
that may affect the outcome of MCT is the exclusion of
many-body correlations~\cite{berthier_critical_2010}. Three-body
static correlations have been shown~\cite{kob_quantitative_2002} to
impact the MCT solutions for a model of silica~\cite{beest__1990}, but
not the ones for the prototypical Kob-Andersen
model~\cite{kob_testing_1995}. Interestingly, a study of a
schematic version of the generalized
MCT~\cite{szamel_2003,wu_cao_2005,mayer_miyazaki_reichman_2006}, which
allows proper description of many-body dynamic correlations, shows
that the ideal transition at $T_\text{MCT}$ can be delayed by
retaining higher order density correlations in the MCT
equations~\cite{mayer_miyazaki_reichman_2006}.

The importance of high order static correlations, hidden in the
amorphous structure of the liquid, is particularly emphasized by
frustration-based approaches to the glass
transition~\cite{kivelson_thermodynamic_1995-1,tarjus_viscous_2000,tanaka_two-order-parameter_2005,tanaka_two-order-parameter_2005-1}.
According to these theories, the phenomenology of glass formation
arises from the competition between the growth of slow, correlated domains,
characterized by some preferred local order, and frustration, which
prevents these domains from percolating through the liquid. Despite
some disagreement on the interpretation of the role of
frustration~\cite{sausset_comment_2008,kawasaki_kawasaki_2008}, these
models indicate medium range order and structural correlations
beyond the pair level as key features for understanding the
dynamic behavior of glass-forming systems.

Computer simulations of several model glassy
systems~\cite{tomida_egami_1995,dzugutov_decoupling_2002,jain_role_2005,shintani_frustration_2006,ladadwa_low-frequency_2006,coslovich_understanding_2007-1,sausset_growing_2010}
and experiments on dense colloidal
suspensions~\cite{patrick_royall_direct_2008,tanaka_critical-like_2010}
provide evidence for the existence of domains formed by preferred
local structures and for their influence on the dynamics. Similar
observations, albeit without explicit reference to the dynamics,
emerge from recent \textit{ab initio} simulations and experiments on
metallic glasses~\cite{fujita_atomic-scale_2009,hirata_direct_2011}.
Furthermore, high order static correlations, named ``point-to-set''
correlations~\cite{montanari_semerjian_2006}, have recently been 
revealed by simulations under amorphous boundary conditions and have
been found to grow by decreasing temperature in a model supercooled
liquid~\cite{biroli_thermodynamic_2008}. In spite of these advances,
there is still no general consensus on the connection between the
structure and dynamics in supercooled liquids. In particular, dynamic
facilitation models~\cite{garrahan_coarse-grained_2003} provide an
alternative and physically appealing description of the glassy
dynamics in terms of purely kinetic constraints.

A clear-cut procedure to test the influence of many-body static
correlations on the dynamics of glass-forming liquids emerges from
recent work of Berthier and
Tarjus~\cite{berthier_nonperturbative_2009,berthier_critical_2010}. These
authors compared the pair structure and dynamics of two model glassy
systems: the Kob-Andersen (KA) binary Lennard-Jones (LJ)
mixture~\cite{kob_testing_1995} and its Weeks-Chandler-Andersen (WCA)
variant~\cite{chandler_lengthscale_2006}, in which the attractive part
of the pair potential is truncated. Berthier and Tarjus found that, at
fixed temperature and for sufficiently large density, the pair structure
of the two models is almost identical, while the structural relaxation
times can differ by orders of magnitude. Thus, direct comparison
of LJ and WCA models offers an ideal benchmark to test the existence
and the influence of static correlations beyond the pair
level. Building on prior knowledge on the preferred local order of LJ
mixtures~\cite{coslovich_understanding_2007-1}, we will consider here
structural indicators of increasing complexity---ranging from pair
correlations, through angular distribution functions, to Voronoi
tessellation---for selected LJ and WCA liquids and perform a crucial
numerical experiment on the influence of structure on the dynamics of
the models.

\section{Methods}

\begin{figure}[!t]
\includegraphics[width=\onefig]{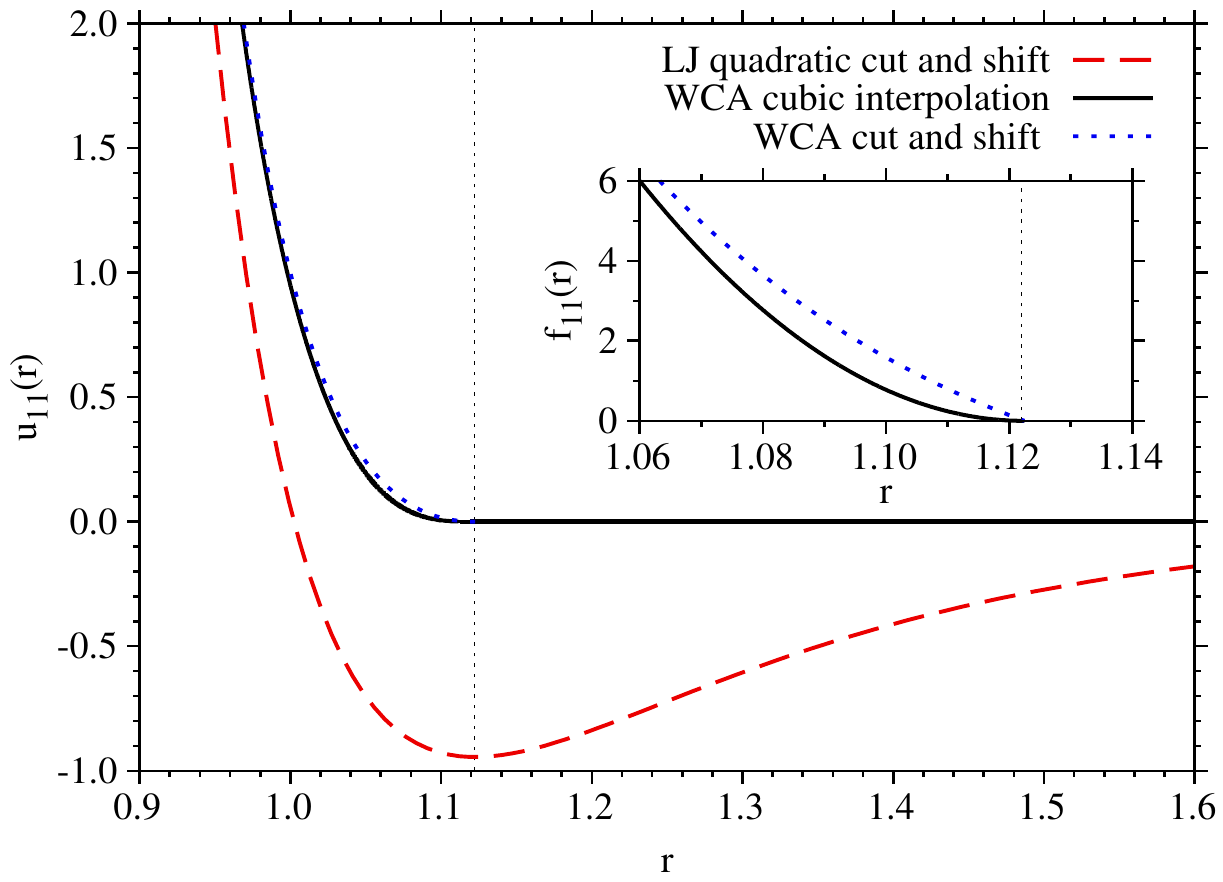}
\caption{\label{fig:u} Main panel: pair potentials between particles
  of species 1 used in this study: LJ (dashed line), WCAS (solid
  line), and WCA (dotted line). Inset: force between particles of
  species 1 for WCAS (solid line) and WCA (dotted line) potentials. In
  both panels, the dotted vertical line marks the distance $r=2^{1/6}$
  corresponding to the minimum of the LJ potential.}
\end{figure}

We will consider two well-known models of glass-forming liquids:
the Kob-Andersen binary mixture~\cite{kob_testing_1995} and the
Wahnstr\"om  binary mixture (WAHN)~\cite{wahnstraem__1991}. In the original
models~\cite{kob_testing_1995,wahnstraem__1991}, named herein KA-LJ
and WAHN-LJ, particles interact through the LJ potential
\begin{equation}\label{eqn:lj} 
u_{\alpha\beta}(r) = 4\epsilon_{\alpha\beta}\left[
  {\left( \frac{\sigma_{\alpha\beta}}{r} \right)}^{12} - {\left(
    \frac{\sigma_{\alpha\beta}}{r} \right)}^6 \right]
\end{equation}
where $\alpha, \beta = 1,2$ are species indices. The values of the
parameters $\sigma_{\alpha\beta}$ and $\epsilon_{\alpha\beta}$ are
$\sigma_{12}=0.8\sigma_{11}$, $\sigma_{22}=0.88\sigma_{11}$,
$\epsilon_{12}=1.5\epsilon_{11}$, and $\epsilon_{22}=0.5\epsilon_{11}$ for
the KA mixture, and $\sigma_{12}=0.916\sigma_{11}$,
$\sigma_{22}=0.833\sigma_{11}$, and
$\epsilon_{22}=\epsilon_{12}=\epsilon_{11}$ for the WAHN mixture. The
chemical compositions and mass rations are $x_1=1-x_2=0.8$,
$m_1/m_2=1$ (for the KA mixture) and $x_1=x_2=0.5$, $m_1/m_2=2$ (for the WAHN mixture).
The potentials are cut and shifted by a quadratic
term~\cite{stoddard_numerical_1973} at $2.5\sigma_{\alpha\beta}$ and
$2.5\sigma_{11}$ in KA and WAHN mixtures, respectively. In the
following, we will use $\sigma_{11}$, $\epsilon_{11}$, and
$\sqrt{m_1\sigma_{11}^2/\epsilon_{11}}$ as units of distance, energy,
and time, respectively. 

In addition, we study the corresponding WCA variants of the above
mixtures. In the WCA
models~\cite{weeks__1971,chandler_lengthscale_2006}, the interaction
parameters $\sigma_{\alpha\beta}$ and $\epsilon_{\alpha\beta}$ and
chemical compositions are unchanged but each of the pair potentials
$u_{\alpha\beta}(r)$ is truncated and shifted so that the value at the
minimum is zero~\cite{weeks__1971}. The WCA truncation of the
attractive part of the potential is well-known from liquid state
theories~\cite{weeks__1971}. We found, however, that this procedure
leads to poor energy conservation during the long molecular dynamics
simulations in the supercooled regime. To circumvent this problem, we
employ here a smooth cut off scheme with cubic
interpolation~\cite{grigera_geometric_2002-1} to ensure continuity up
to the second derivative of the potentials at the minimum
$r_c=2^{1/6}\sigma_{\alpha\beta}$ of the LJ potential. Explicitly, the
WCA smoothed potentials (WCAS) read
\begin{equation}
u^s_{\alpha\beta}(r) = \left\{
\begin{array}{ll}
u_{\alpha\beta}(r) + A_{\alpha\beta} & r<a_{\alpha\beta}\\
B_{\alpha\beta} (r_c-r)^3 & a_{\alpha\beta}<r<r_c \\
0 & r>r_c
\end{array}
\right.
\end{equation}
where $A_{\alpha\beta}$ and $B_{\alpha\beta}$ are determined to ensure
continuity at $r=a$ and $r=r_c$. The parameters $a_{\alpha\beta}$ are
adjusted for each pair $\alpha$-$\beta$ so that
$r_c=2^{1/6}\sigma_{\alpha\beta}$, and read $a_{11}=1.0269$,
$a_{12}=0.8215$, $a_{22}=0.9038$ for the KA-WCAS mixture, and
$a_{11}=1.0269$, $a_{12}=0.9473$, $a_{22}=0.8555$ for the WAHN-WCAS
mixture. A comparison between LJ, WCA, and WCAS potentials for 1-1
pairs is shown in Fig.~\ref{fig:u}. In contrast to the WCA potential,
the derivative of the force of the WCAS potential is continuous at
$r_c$. In the inset, we highlight the difference
between the WCA and the WCAS potentials in the $r\sim r_c$ region. As it will
be clear in the following, this modification introduces some small
differences in the thermodynamic and dynamic properties, but does not
qualitatively alter the comparison with LJ models. In the following,
we will mostly concentrate on the WCAS models and report selected
results for the original WCA models. All studied systems are composed
of 1000 particles in a cubic box with periodic boundary
conditions. Molecular dynamics simulations are performed in the NVT
ensemble using the Nos\'e-Poincar\'e thermostat~\cite{nose__2001} with
a mass parameter $Q$=5.0. The number density of the KA mixtures is
$\rho=1.2$, while that of Wahnstr\"om mixtures is $\rho=1.297$. For
the LJ and WCAS models, static and dynamic properties are averaged over up to six
independent thermal histories. We find that the KA-WCAS mixture
crystallizes more easily than the other systems~\footnote{In this
  case, 4 samples out 6 crystallized around $T\sim 0.3$.}. A similar
tendency to crystallize has been reported in
Ref.~\cite{toxvaerd_stability_2009} for the KA-WCA model. Only the
non-crystallizing samples are retained to perform the averages.

\section{Results}

\subsection{Two-body and three-body static correlations}

\begin{figure}[!t]
\includegraphics[width=\onefig]{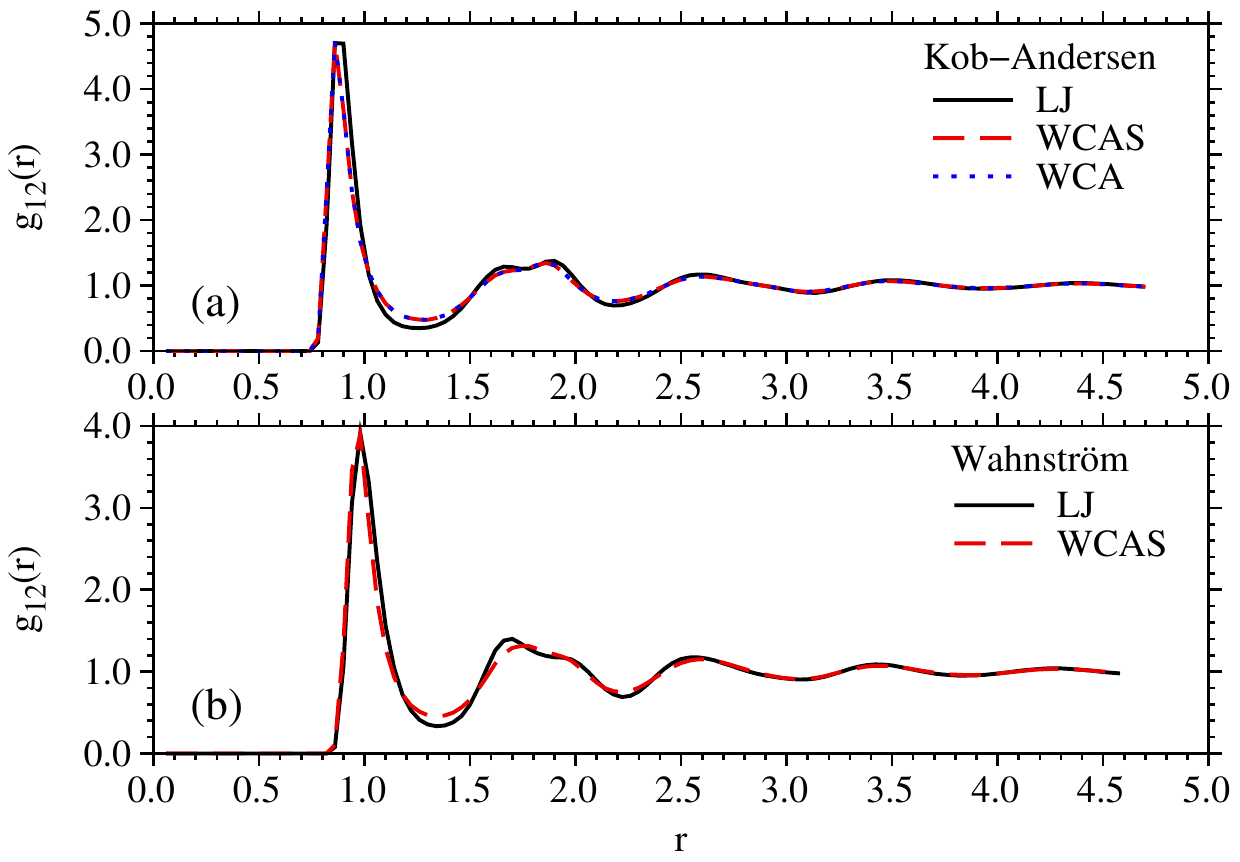}
\caption{\label{fig:gr} Radial distribution functions $g_{12}(r)$ for
  (a) the Kob-Andersen mixture and (b) the Wahnstr\"om mixture. The state
  points considered are (a) $\rho=1.2$, $T=0.5$ and (b) $\rho=1.297$,
  $T=0.6$. In both panels, solid, dashed, and dotted lines indicate
  results for the LJ, WCAS, and WCA models, respectively. Error bars
  are smaller than the widths of the lines.}
\end{figure}

To start the discussion, we analyze the pair structure of the present
models. Figure~\ref{fig:gr} displays the radial distribution function
$g_{12}(r)$ between unlike species in the KA mixtures (top panel) and WAHN
mixtures (bottom panel). For each type of mixture, the results obtained
are shown at a common temperature, representative of the slow dynamics
regime of the LJ models. The pair structure appears essentially
unaffected by the truncation of attractions, thus confirming the
observations of recent simulation
works~\cite{berthier_critical_2010,pedersen_repulsive_2010}. This
result holds for both WCA and WCAS models. Only a close inspection of
the figures reveals that the first minima of $g_{12}(r)$ are slightly
deeper in the LJ models, suggesting that the latter systems are
effectively more supercooled. A similar effect is visible in the
radial distribution functions reported by Pedersen
\etal~\cite{pedersen_repulsive_2010} for KA-LJ and KA-WCA mixtures.

\begin{figure}[!t]
\includegraphics[width=\onefig]{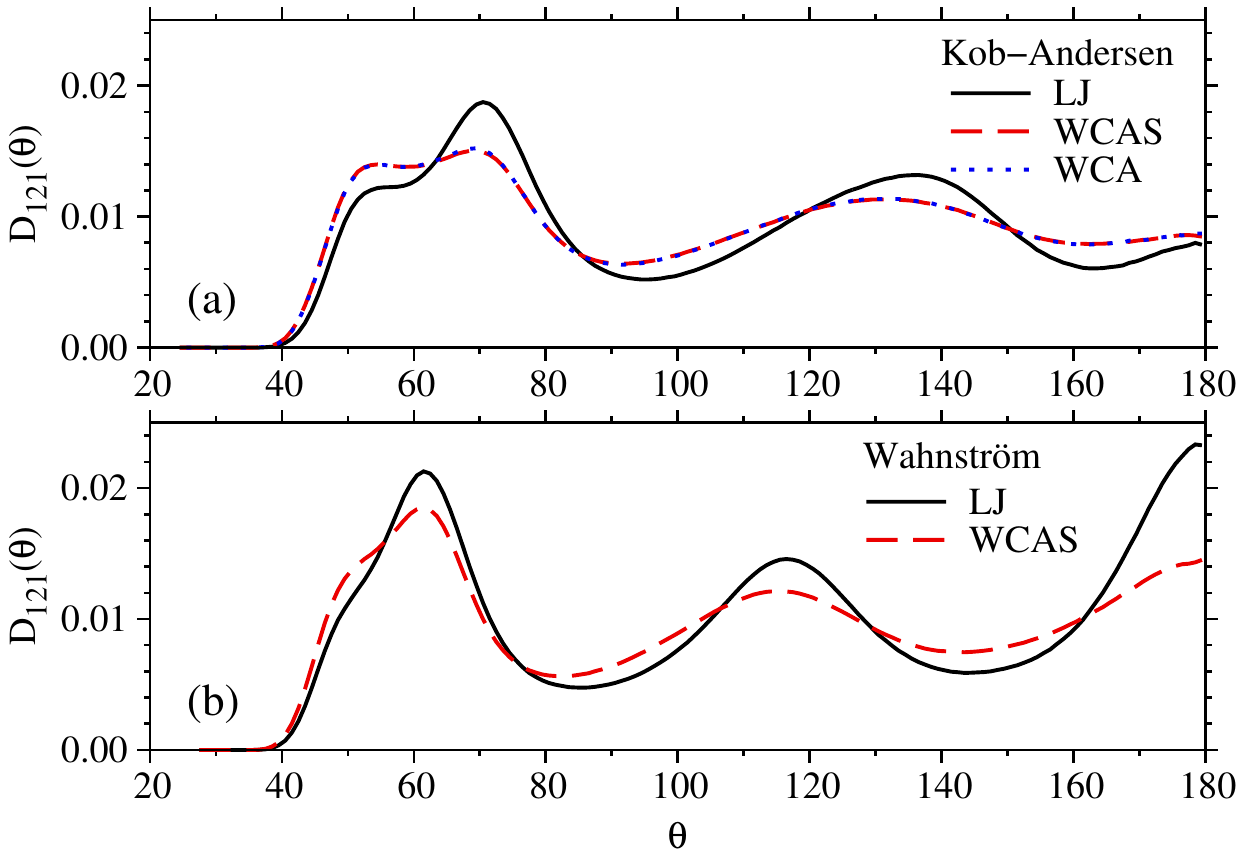}
\caption{\label{fig:angle1} Angular distribution functions
  $D_{121}(r)$ for (a) Kob-Andersen mixtures and (b) Wahnstr\"om
  mixtures. State points and lines are the same as in Fig.~\ref{fig:gr}. Error bars
  are smaller than the widths of the lines.}
\end{figure}

\begin{figure}[!t]
\includegraphics[width=\onefig]{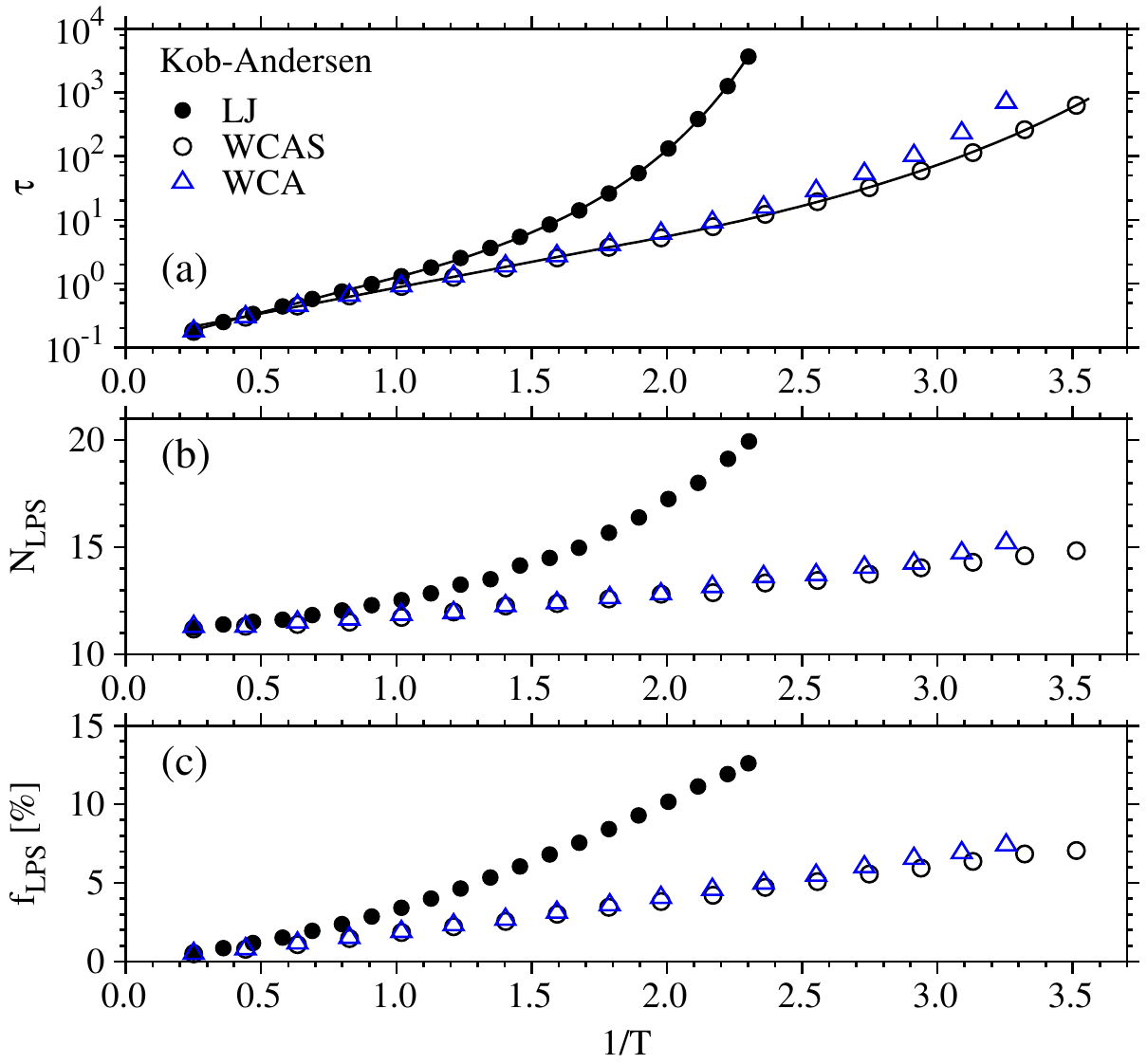}
\caption{\label{fig:lps1} (a) Structural relaxation times $\tau$ as a
  function of $1/T$ for Kob-Andersen mixtures: LJ (filled circles),
  WCAS (open circles), and WCA (open triangles) models. The
  wave-vector considered for the calculation of $\tau$ is
  $k=7.0$. Fits to the modified VFT equation (Eq.~\eqref{eq:vft}) are
  shown as solid lines. (b) Average number $N_\lps$ of particles in
  LPS domains formed by (0,2,8)-polyhedra. Symbols have the same
  meaning as in (a). (c) Average fraction of particles of species 2 at
  the center of (0,2,8) polyhedra as a function of $1/T$. Symbols have
  the same meaning as in (a).}
\end{figure}

At the temperatures considered in Fig.~\ref{fig:gr}, the structural
relaxation times of the LJ and WCA models differ by almost two
orders of magnitude, as is evident from
Ref.~\cite{berthier_critical_2010} and Figs.~\ref{fig:lps1}
and~\ref{fig:lps2} (discussed in further detail below). Given the small
differences observed in the pair structure, it is natural to ask
whether this large variation in the dynamic properties is due to
static correlations of higher order. In a first attempt to go beyond pair
correlations, we calculate the bond-angle distribution functions
$D_{\alpha\beta\gamma}(\theta)$ between triplets of neighboring
particles of particles of species $\alpha$, $\beta$, and $\gamma$,
where $\beta$ is the species of the central
particle. Figure~\ref{fig:angle1} shows the angular distribution
functions $D_{121}(\theta)$ for the same state points considered in
Fig.~\ref{fig:gr}. Angular correlations reveal more clearly the
structural differences between the LJ and WCA models. In KA mixtures, the
sharp peaks in $D_{121}(\theta)$ around $\sim 70^\circ$ and the broad
peak in the range $120^\circ - 140^\circ$ reflect local arrangements
corresponding to distorted twisted bi-capped prisms of large particles
(species 1) centered around small particles (species
2)~\cite{coslovich_understanding_2007-1}. A comparison of the LJ and WCA
data sets thus reveals that the KA-LJ mixture has a more pronounced
local ordering than the KA-WCA mixture at the selected thermodynamic state. As
in the case of $g_{12}(r)$, the difference between the WCA and WCAS models
is negligible for this state point. A similar effect is visible for
the WAHN mixture: the peaks in $D_{121}(\theta)$, located around
$63^\circ$, $116^\circ$, and $180^\circ$, are signatures of local
icosahedral ordering, which appears more pronounced in the original LJ
model than in the WCAS variant. Similar conclusions can be drawn from an
analysis of the other angular distribution functions (not shown here)
and are corroborated by an inspection of data at even lower
temperature. We conclude that the structure of LJ and WCA systems
differ more evidently at the level of three-body static correlations
and that the increase of local ordering upon switching on attractions
correlate qualitatively with the increase of relaxation times.

\subsection{Locally preferred structures}

\begin{table*}[h]
  \caption{\label{tab1}Most frequent signatures of Voronoi polyhedra around
    particles of species 2 from instantaneous configurations and local
    minima of potential energy surface. Percentages are calculated
    with respect to the number of particles of species 2. The low temperature data set
    ($T=T_l$) corresponds to the lowest available temperatures:
    $T=0.435$ (KA-LJ), $T=0.285$ (KA-WCAS), $T=0.560$ (WAHN-LJ), and 
    $T=0.343$ (WAHN-WCAS). The high temperature data set ($T\approx
    T^*$) corresponds to temperatures close to the crossover temperature
    $T^*$: $T=0.983$ (KA-LJ), $T=0.627$ (KA-WCAS), $T=1.072$ (WAHN-LJ), and
    $T=0.598$ (WAHN-WCAS).}
\begin{tabular*}{\linewidth}{@{\extracolsep{\fill}} l rl rl rl rl}
\hline
\hline
         & \multicolumn{4}{c}{Instantaneous configurations} & \multicolumn{4}{c}{Local minima} \\
\cline{2-5} \cline{6-9}
         & \multicolumn{2}{c}{$T=T_l$} & \multicolumn{2}{c}{$T\approx T^*$} & \multicolumn{2}{c}{$T=T_l$} & \multicolumn{2}{c}{$T\approx T^*$} \\
\cline{2-3} \cline{4-5} \cline{6-7} \cline{8-9}
         & \% & Signature & \% & Signature &\% & Signature &\% & Signature \\
\hline
KA-LJ &      	        12.6 &      (0,2,8) & 	         4.0 &    (0,2,8,1) & 	        19.9 &      (0,2,8) & 	        12.4 &      (0,2,8) \\
         &      	         8.3 &    (1,2,5,3) & 	         3.9 &    (1,2,5,3) & 	         7.2 &    (1,2,5,3) & 	         5.9 &    (1,2,5,3) \\
         &      	         5.6 &    (1,2,5,2) & 	         3.4 &      (0,2,8) & 	         6.8 &    (1,2,5,2) & 	         5.8 &    (1,2,5,2) \\
         &      	         5.1 &      (0,3,6) & 	         3.2 &    (0,4,4,3) & 	         6.7 &      (0,3,6) & 	         5.3 &    (0,3,6,1) \\
\hline
KA-WCAS &      	         7.1 &      (0,2,8) & 	         3.9 &    (0,2,8,1) & 	         8.4 &      (0,2,8) & 	         5.2 &      (0,2,8) \\
         &      	         5.7 &    (1,2,5,3) & 	         3.5 &    (1,2,5,3) & 	         5.4 &    (1,2,5,3) & 	         4.1 &    (1,2,5,3) \\
         &      	         4.7 &    (0,2,8,1) & 	         3.4 &    (0,3,6,3) & 	         4.5 &    (1,2,5,2) & 	         4.0 &    (0,4,4,3) \\
         &      	         3.7 &    (0,4,4,3) & 	         3.1 &    (0,4,4,3) & 	         4.3 &    (0,2,8,1) & 	         3.9 &    (1,2,5,2) \\
\hline
WAHN-LJ &      	        27.4 &     (0,0,12) & 	         7.0 &    (0,3,6,4) & 	        32.7 &     (0,0,12) & 	        10.9 &     (0,0,12) \\
         &      	         9.0 &    (0,2,8,2) & 	         5.2 &    (0,2,8,2) & 	        10.0 &    (0,2,8,2) & 	        10.6 &    (0,2,8,2) \\
         &      	         7.7 &   (0,1,10,2) & 	         3.5 &   (0,1,10,2) & 	         8.3 &   (0,1,10,2) & 	         9.7 &    (0,3,6,4) \\
         &      	         6.0 &    (0,3,6,4) & 	         3.0 &    (0,3,6,3) & 	         6.5 &    (0,3,6,4) & 	         7.3 &   (0,1,10,2) \\
\hline
WAHN-WCAS &      	        18.9 &     (0,0,12) & 	         7.7 &    (0,3,6,4) & 	        20.0 &     (0,0,12) & 	         8.5 &    (0,3,6,4) \\
         &      	         9.1 &    (0,2,8,2) & 	         6.4 &    (0,2,8,2) & 	         9.8 &    (0,2,8,2) & 	         7.9 &    (0,2,8,2) \\
         &      	         7.2 &    (0,3,6,4) & 	         4.3 &   (0,1,10,2) & 	         7.4 &    (0,3,6,4) & 	         5.8 &     (0,0,12) \\
         &      	         6.9 &   (0,1,10,2) & 	         4.1 &     (0,0,12) & 	         6.6 &   (0,1,10,2) & 	         4.7 &   (0,1,10,2) \\
\hline
\hline
\end{tabular*}
\end{table*}

\begin{table*}[h]
  \caption{\label{tab2}Same as Table~\ref{tab1} but for Voronoi polyhedra around
    particles of species 1. Percentages are calculated
    with respect to the number of particles of species 1.}
\begin{tabular*}{\linewidth}{@{\extracolsep{\fill}} l rl rl rl rl}
\hline
\hline
         & \multicolumn{4}{c}{Instantaneous configurations} & \multicolumn{4}{c}{Local minima} \\
\cline{2-5} \cline{6-9}
         & \multicolumn{2}{c}{$T=T_l$} & \multicolumn{2}{c}{$T\approx T^*$} & \multicolumn{2}{c}{$T=T_l$} & \multicolumn{2}{c}{$T\approx T^*$} \\
\cline{2-3} \cline{4-5} \cline{6-7} \cline{8-9}
         & \% & Signature & \% & Signature &\% & Signature &\% & Signature \\
\hline
KA-LJ &      	         7.8 &    (0,2,8,4) & 	         4.7 &    (0,2,8,4) & 	         8.7 &    (0,2,8,4) & 	         7.2 &    (0,2,8,4) \\
         &      	         5.7 &    (0,2,8,5) & 	         3.2 &    (0,3,6,5) & 	         6.6 &    (0,2,8,5) & 	         5.1 &    (0,2,8,5) \\
         &      	         5.0 &    (0,3,6,6) & 	         3.0 &    (0,3,6,4) & 	         5.3 &    (0,3,6,6) & 	         4.1 &    (0,3,6,6) \\
         &      	         4.4 &    (0,3,6,5) & 	         2.9 &    (0,2,8,5) & 	         4.8 &   (0,1,10,4) & 	         4.1 &   (0,1,10,2) \\
\hline
KA-WCAS &      	         7.4 &    (0,2,8,4) & 	         5.2 &    (0,2,8,4) & 	         7.5 &    (0,2,8,4) & 	         6.1 &    (0,2,8,4) \\
         &      	         5.0 &    (0,2,8,5) & 	         3.5 &    (0,3,6,5) & 	         5.1 &    (0,2,8,5) & 	         3.9 &    (0,2,8,5) \\
         &      	         4.7 &    (0,3,6,6) & 	         3.2 &    (0,3,6,4) & 	         4.9 &    (0,3,6,6) & 	         3.8 &    (0,3,6,5) \\
         &      	         4.5 &    (0,3,6,5) & 	         3.2 &    (0,3,6,6) & 	         4.4 &    (0,3,6,5) & 	         3.8 &    (0,3,6,6) \\
\hline
WAHN-LJ &      	         8.0 &   (0,1,10,4) & 	         4.7 &    (0,2,8,4) & 	         9.5 &   (0,1,10,4) & 	         7.4 &    (0,2,8,5) \\
         &      	         7.1 &    (0,2,8,5) & 	         3.7 &    (0,2,8,5) & 	         8.5 &    (0,2,8,5) & 	         6.8 &    (0,2,8,4) \\
         &      	         6.3 &    (0,2,8,4) & 	         3.2 &    (0,3,6,6) & 	         6.7 &    (0,2,8,4) & 	         5.4 &   (0,1,10,4) \\
         &      	         4.7 &   (0,1,10,3) & 	         3.0 &    (0,3,6,5) & 	         5.2 &   (0,1,10,3) & 	         4.6 &    (0,3,6,6) \\
\hline
WAHN-WCAS &      	         7.6 &    (0,2,8,5) & 	         5.8 &    (0,2,8,4) & 	         8.0 &    (0,2,8,5) & 	         6.3 &    (0,2,8,4) \\
         &      	         6.9 &    (0,2,8,4) & 	         4.8 &    (0,2,8,5) & 	         6.9 &    (0,2,8,4) & 	         5.8 &    (0,2,8,5) \\
         &      	         6.8 &   (0,1,10,4) & 	         3.8 &    (0,3,6,6) & 	         6.8 &   (0,1,10,4) & 	         4.4 &    (0,3,6,6) \\
         &      	         4.6 &   (0,1,10,3) & 	         3.4 &    (0,3,6,5) & 	         4.5 &   (0,1,10,3) & 	         3.6 &   (0,1,10,4) \\
\hline
\hline
\end{tabular*}

\end{table*}

To render the connection between the local structure and dynamics
explicit, we analyze the statistics of Voronoi polyhedra as a function
of temperature. The Voronoi tessellation implicitly entails more
complex static correlations (although, of course, it cannot be
expressed as a multi-particle correlation function) and reveals the
details of the particles' arrangements within the first coordination
shell. Inspection of the spatial persistence of a given local
structure provides information on extended structural correlations,
\textit{i.e.}, medium range order. The protocol adopted here is the same as
that used in a prior investigation of the local structure of binary LJ mixtures
at constant pressure~\cite{coslovich_understanding_2007-1}. We monitor
the temperature dependence of the fraction of Voronoi polyhedra with a
given signature $(n_3, n_4, \dots)$, where $n_i$ is the number of
faces of the polyhedron with a given number $i$ of vertices. We
identify the locally preferred structure (LPS) of the liquid as the
geometrical arrangement corresponding to the most frequent Voronoi
polyhedron around particles of species 2 observed in the samples at
low temperature. This choice is based on the observation that in
binary LJ mixtures it is easier to characterize local order around
small
particles~\cite{valle__1994,coslovich_understanding_2007-1}. This is
also consistent with a previous study of the KA-LJ mixture, which
focused on the coordination polyhedra of large particles around the small
ones~\cite{Fernandez_Harrowell_2004}. The most frequent signatures of
Voronoi polyhedra around particles of species 2 and 1 are reported in
Table~\ref{tab1} and~\ref{tab2}, respectively. We find that the
typical Voronoi polyhedra observed around particles of species 1 [such
as (0,2,8,4) or (0,1,10,4) polyhedra] do not display evident
symmetries and lack a clear structural identification. Furthermore,
the corresponding percentages do not increase as sharply upon
decreasing temperature as those calculated for Voronoi polyhedra
around particles of species 2. In the following, we will therefore
base our analysis on the local structures observed around this latter
type of particles. Understanding the nature of local order around
large particles remains an open issue, which may need more refined
methods to detect short and medium range order.

\begin{figure}[!t]
\includegraphics[width=\onefig]{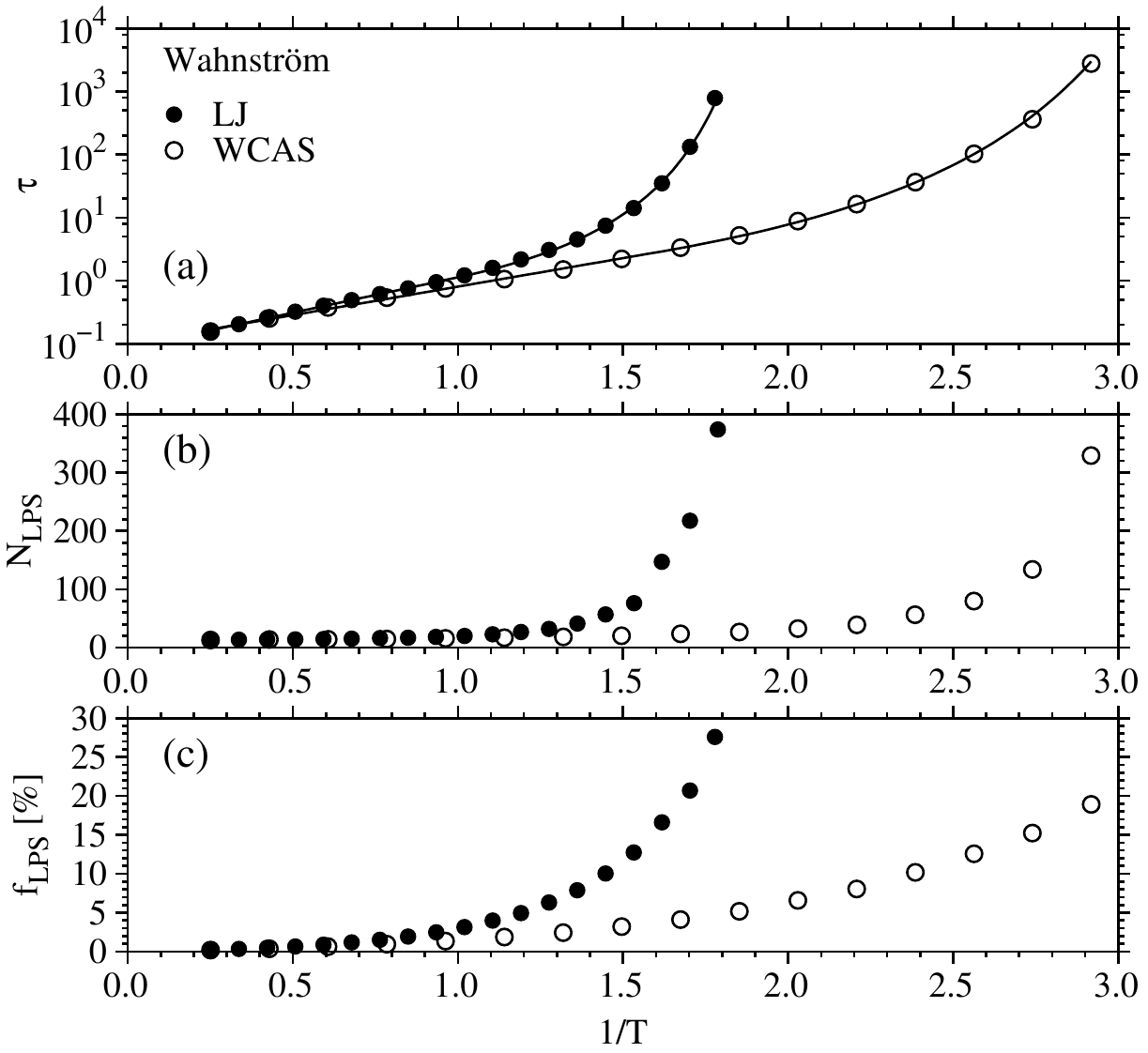}
\caption{\label{fig:lps2} Same as Fig.~\ref{fig:lps1} but for 
  Wahnstr\"om mixtures. For these systems, the LPS corresponds to
  (0,0,12) polyhedra.}
\end{figure}

Our results confirm the observations of
Ref.~\cite{coslovich_understanding_2007-1} at constant pressure: in
the low temperature regime, (0,2,8) and (0,0,12) polyhedra around
particles of species 2 constitute the dominant signatures in KA-LJ and
WAHN-LJ mixtures, respectively. Thus, we identify the LPS of KA-LJ and
WAHN-LJ as twisted bi-capped square prisms and icosahedra,
respectively~\cite{coslovich_understanding_2007-1}. The identification
is consistent with previous investigations on KA-LJ
clusters~\cite{doye_controlling_2007} and with a recent simulation
study on the bulk WAHN-LJ
mixture~\cite{pedersen_geometry_2010}. By applying the same procedure to
the WCA and WCAS models, we find that the locally preferred structures
remain the same as in the original LJ models. We observe, however, a
systematic reduction of the fraction of LPS upon truncating the
attractions. This effect will be discussed in further detail below.

As a general rule, the fraction $f_\lps$ of particles of species 2
at the center of a LPS increases with decreasing
temperature~\cite{coslovich_understanding_2007-1}. The growth of
$f_\lps$ reflects the formation of slow, long-lived clusters of
neighboring LPS~\cite{coslovich_understanding_2007-1}. A calculation of
the self intermediate scattering functions filtered according to the
pertinent Voronoi polyhedra, shows that the typical relaxation times
of particles at the center of LPSs are up to 10 times larger than those
outside LPSs~\cite{coslovich_understanding_2007-1}. In the following, we
will refer to these clusters as ``LPS domains'', which are defined as
groups of particles sitting either at the center or on the vertices of
face-sharing polyhedra with the signature of the LPS. The average
number of particles forming an LPS domain will be denoted by
$N_\lps$, which is a measure of the spatial extension over which the
liquid adopts the same preferred local structure~\footnote{Note that
  this correlation does not require neighboring LPS to be
  \textit{oriented} in a similar fashion, provided a preferred
  direction of the LPS could be defined.}.  In Figs.~\ref{fig:lps1}(b) and~\ref{fig:lps2}(b) we show $N_\lps$ as a function
of $1/T$ for the KA and WAHN models, respectively. To facilitate
a comparison with previous work~\cite{coslovich_understanding_2007-1}, we
include the temperature dependence of $f_\lps$ in Figs.~\ref{fig:lps1}(c) and~\ref{fig:lps2}(c). 
Note that, while $f_\lps$ is evaluated with respect to
particles of species 2, both species of particles contribute to the
size $N_\lps$ of LPS domains. Both $N_\lps$ and $f_\lps$ increase in a similar fashion as
$T$ decreases, although with slightly different functional forms. The
growth of LPS domains is particularly dramatic in WAHN mixtures, which
develop a strong icosahedral order upon supercooling. By contrast, the
size of the domains formed by prismatic structures in KA models is
relatively small (20--30 particles). Nonetheless, the structural
evolution in all the systems studied follow qualitatively similar
patterns.

\subsection{Connection between structure and dynamics}

To illustrate the connection to the dynamics of the models, we now
study the temperature dependence of the structural relaxation times
$\tau$. The latter are defined by the condition $F_s(k=7,\tau)=1/e$,
where $F_s(k,t)$ is the self intermediate scattering function
averaged over all particles. The relaxation times have been fitted by
the following modified Vogel-Fulcher-Tammann (VFT)
equation~\cite{coslovich_understanding_2007-1}
\begin{equation}\label{eq:vft}
\tau(T) = 
\left\{ 
\begin{array}{ll}
  \tau_{\infty}     \exp\left[E_{\infty}/T\right] & T>T^*   \\
  \tau_{\infty}^{'} \exp\left[\dfrac{1}{K(T/T_0-1)}\right] & T<T^* \\
\end{array}
\right.
\end{equation}
where
\begin{equation}
\tau_{\infty}^{'}=\tau_{\infty}
\exp\left[E_{\infty}/T^{*} - \dfrac{1}{K(T^{*}/T_0-1)}\right]
\end{equation}
Equation~\eqref{eq:vft} ensures a smooth crossover around $T^*$
between the Arrhenius law at high $T$ and the VFT equation
at low $T$, accounting for the super-Arrhenius dependence of the relaxation
times.
Figures~\ref{fig:lps1}(a) and \ref{fig:lps2}(a) display
$\tau$ as a function of $1/T$ for the KA and WAHN mixtures, respectively,
together with the corresponding fits to Eq.~\eqref{eq:vft}. Figure~\ref{fig:lps1}(a) also includes results for the KA-WCA
mixture obtained using the original cut and shift at the minimum of
the potentials, as in previous
works~\cite{chandler_lengthscale_2006,berthier_nonperturbative_2009,berthier_critical_2010}. The
latter data set is in good agreement with the results obtained in
Refs.~\cite{berthier_nonperturbative_2009,berthier_critical_2010} for a
similar wave-vector ($k=7.21$). At sufficiently low temperature,
however, non-negligible deviations appear with respect to the KA-WCAS
mixture. This discrepancy may be attributed to the modification
induced by the smooth cut off employed in this work. The comparison
between the LJ and WCAS models, however, remains qualitatively unaffected
and confirms the general conclusions of
Ref.~\cite{berthier_nonperturbative_2009,berthier_critical_2010}.

A comparison of the LPS analysis and relaxation times data reveals a
striking relationship between structure and dynamics. The increase of
$\tau$ below the crossover temperature $T^*$ correlates to the
increase in size of the LPS domains. This connection is particularly
evident for the two WAHN mixtures, in which the increase of
icosahedral order upon decreasing $T$ is very sharp. Our results
reveal that the large difference in the dynamic behavior between the LJ and
WCA models reflects different stages of the evolution of the local
structure on the way to glass formation---an effect that is barely
visible at the level of pair correlations. It should also be noted that the size of
the LPS domains at low temperature is slightly larger in the WCA than in the WCAS
models, which is consistent with the discrepancy in relaxation times
observed above.

A remark on the nature of local order in these models is in order. It
has been shown recently that the WAHN-LJ mixture can phase-separate
and then partially crystallize in a complex crystal structure that
accommodates distorted icosahedral
geometries~\cite{pedersen_geometry_2010}. The LPS observed in the
liquid is thus analogous to the typical local structure of the
underlying crystal, which is at odds with the paradigm of the
frustration-limited domains theory~\cite{tarjus_viscous_2000}. The
situation is less clear in the case of the KA-LJ mixture, for which an
unambiguous identification of the crystalline phase is
lacking. Previous
studies~\cite{fernandez_organization_2004,valdes_mixing_2009} have
shown that, for chemical compositions close to the one of the original
model, stable crystals either have CsCl symmetry or are
composed of a mixture of fcc and hcp structures of large
particles. Interestingly, we found that the KA-WCAS model can
crystallize into a fcc lattice of large particles. In
this model crystallization is associated with a sudden drop in the fraction of (0,2,8)
polyhedra and a rapid increase of (0,4,4,6) polyhedra centered around
particles of species 1. This reveals a potential mismatch between the
LPS of the liquid and the typical local structure of the crystal. The
question of whether the locally preferred structure should coincide with the structure of the underlying
crystal~\cite{sausset_comment_2008,kawasaki_kawasaki_2008} certainly
deserves further investigations.

\begin{figure}[!t]
\includegraphics[width=\onefig]{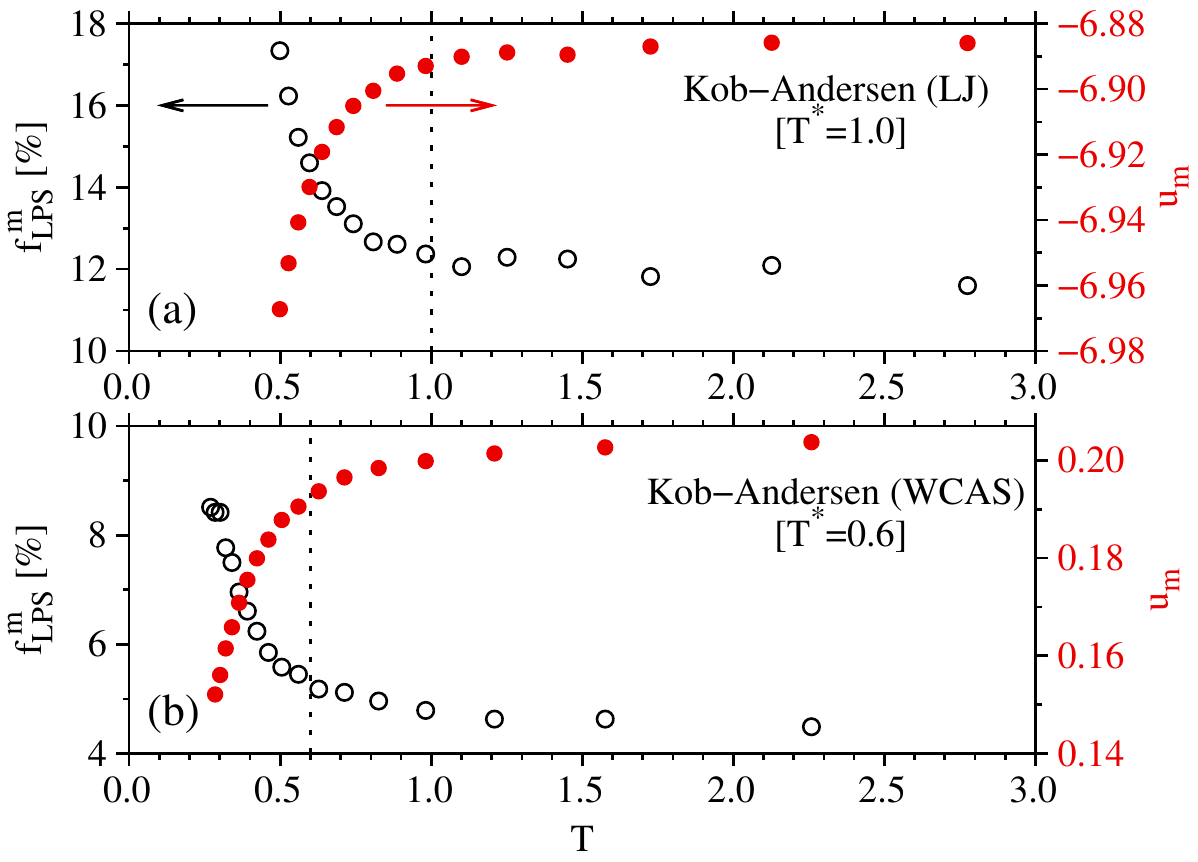}
\caption{\label{fig:onset1} Average fraction $f^m_\lps$ of LPS in
  local minima (open circles, left axis) and average potential energy
  $u_m$ of local minima (filled circles, right axis) as a function of
  $T$ for Kob-Andersen mixtures: (a) LJ model and (b) WCAS model. The
  dynamic crossover temperatures $T^*$ obtained from fits to
  Eq.~\eqref{eq:vft} are indicated as vertical dotted lines.}
\end{figure}

\begin{figure}[!h]
\includegraphics[width=\onefig]{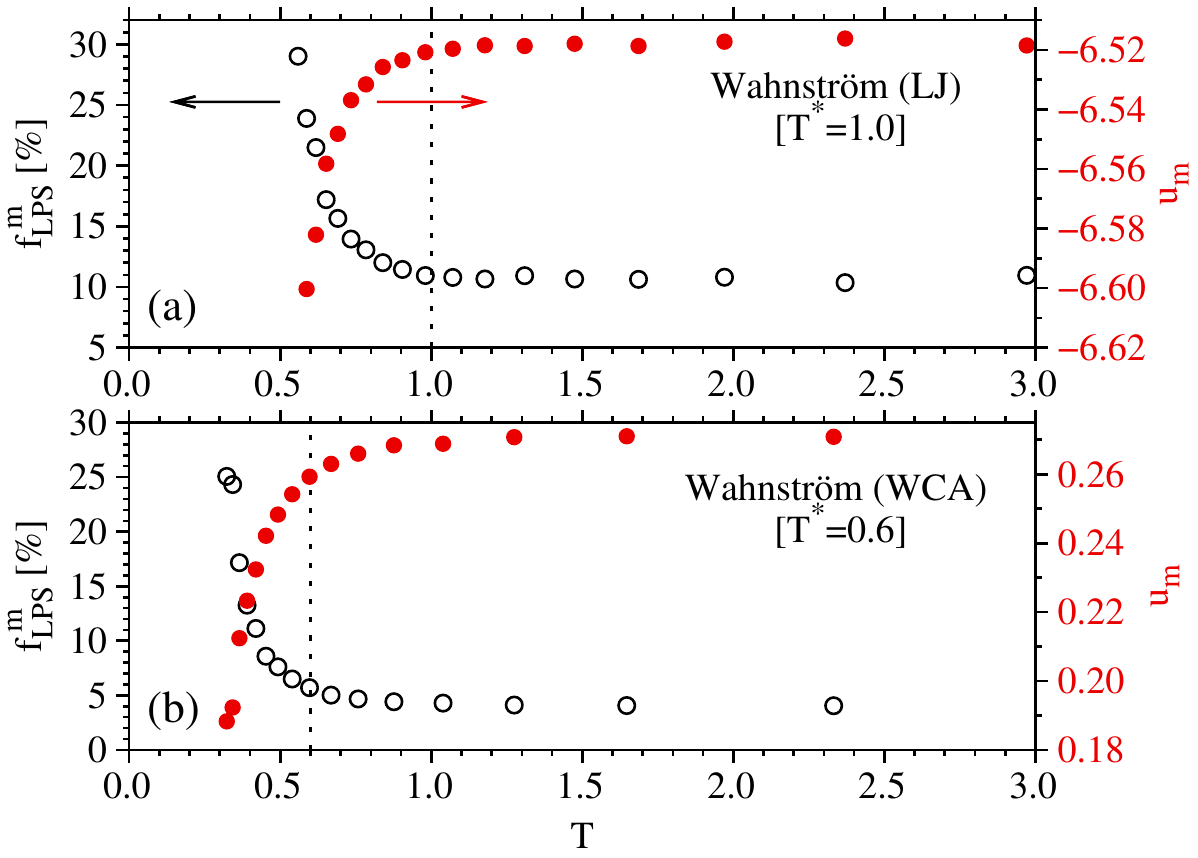}
\caption{\label{fig:onset2} Same as Fig.~\ref{fig:onset1} but for 
  Wahnstr\"om mixtures.}
\end{figure}

\subsection{Connections between structure, potential energy landscape
  and fragility}

The connection between the growth of LPS domains and slow dynamics is
further corroborated by the analysis of the potential energy landscape
(PEL). It is well-known that the appearance of super-Arrhenius
behavior and non-exponential relaxation around the so-called onset
temperature $T_O$ coincides with a sharp change in the properties of
the local minima of the PEL explored by system~\cite{sastry__1998}. In fact,
the average energy $u_m(T)$ of local minima remains almost
constant at high $T$ and starts decreasing rapidly below
$T_O$. Figures~\ref{fig:onset1} and~\ref{fig:onset2} display $u_m$ and
the fraction $f^m_\lps$ of particles of species 2 at the center of the LPSs
evaluated for local minima of the PEL as a function of temperature for
the KA and WAHN mixtures, respectively. We note that the values of $T^*$
obtained from Eq.~\ref{eq:vft} are only slightly larger than the onset
temperatures estimated from the appearance of two-step,
non-exponential relaxation in the dynamic correlation functions, and
are consistent, at least for KA mixtures, with the values of $T_O$
reported in Ref.~\cite{elmatad_chandler_garrahan_2010}.

Strikingly, the onset of slow dynamics, indicated by the drop in
$u_m$, always correlates to a sharp increase of $f^m_\lps$. We remark
that the percentages of other, less frequent signatures of Voronoi
polyhedra do not increase as sharply across $T^*$, although some of
them do display some change upon decreasing $T$ (see
Table~\ref{tab1}). We thus attribute the onset of slow dynamics in the
models studied herein to the growth of structural correlations. We also note
that landscape approaches based on high-order stationary points of the
potential energy surface~\cite{grigera_geometric_2002} may provide a
complementary view on our results. We found in fact that particles at
the center of LPSs participate less to \textit{unstable} modes of
saddle
points~\cite{coslovich_phd,coslovich_understanding_2007,coslovich_unpublished}. Establishing
a clear relationship between the growth of LPSs and the disappearance
of unstable directions in the landscape is an interesting open issue
that is left for future investigation.

The present findings ostensibly indicate that the drop in $u_m$ is connected
to the formation of energetically favored structures. This is
consistent with the observation that the potential energy associated
with the LPS is \textit{typically} lower than that of other
structures~\cite{coslovich_phd}. In general, however, specific local
structures may be favored also for non-energetic reasons, such as more
efficient packing or non-trivial entropic effects (e.g., favorable
arrangements of interlocking LPS). An interesting example of this
competition is provided by a model of a NiY alloy based on LJ
interactions~\cite{valle__1994}, for which the LPS---a capped trigonal
prismatic structure---corresponds to a Voronoi polyhedron having the
smallest volume but not the lowest potential
energy~\cite{coslovich_phd}. Successful attempts to account for this
complex interplay at the mean-field level in the one-component LJ
liquid~\cite{mossa_locally_2003} and in a soft-sphere
mixture~\cite{hentschel_statistical_2007,lerner_statistical_2009} must
also be acknowledged. 

\begin{figure}[t]
\includegraphics[width=\onefig]{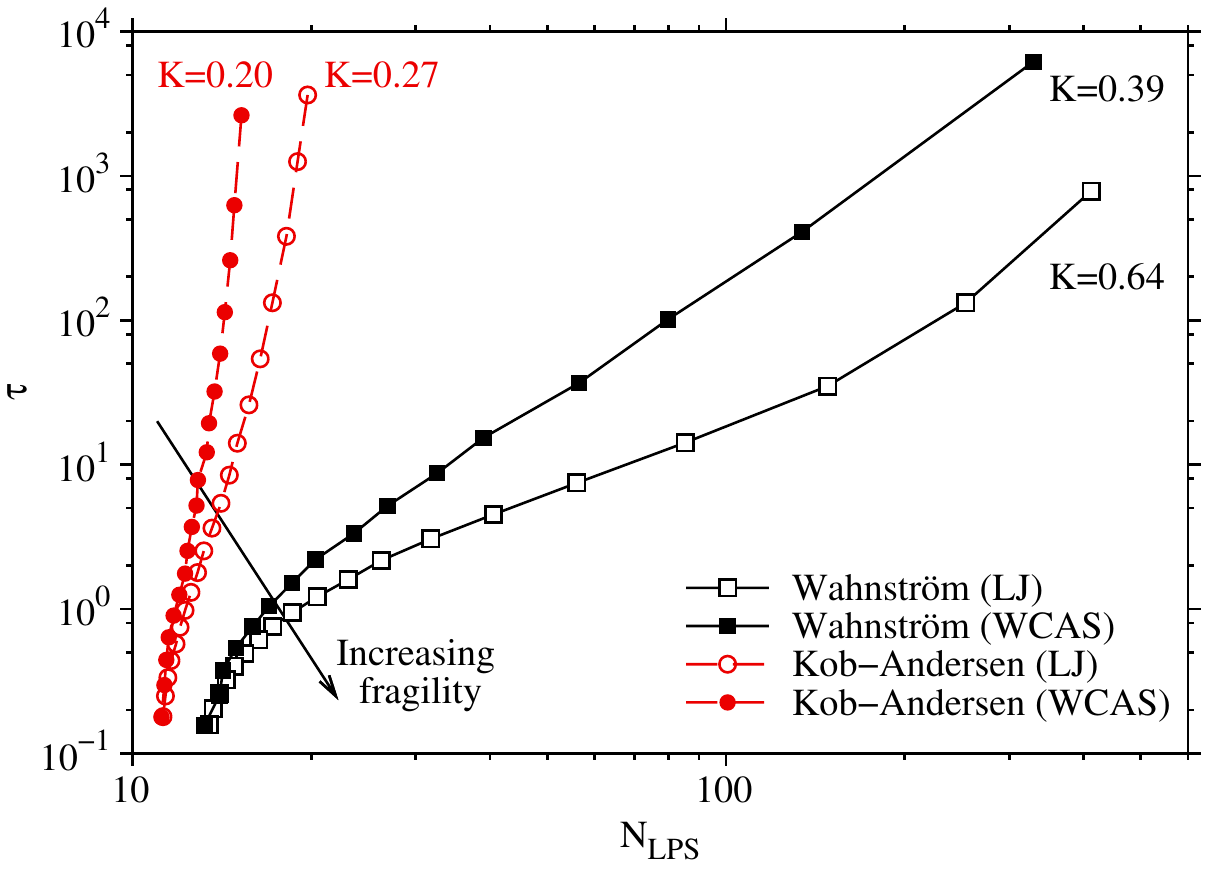}
\caption{\label{fig:lps_vs_tau}Relaxation times $\tau$ as a function
  of the average number $N_\lps$ of particles in the LPS domains. The
  corresponding fragility indices $K$ of the models, obtained from
  fits to the modified VFT equation [Eq.~\eqref{eq:vft}], are also
  indicated.}
\end{figure}

To set forth the present results in a more compact fashion, we now plot $\tau$
as a function of $N_\lps$ (see Fig.~\ref{fig:lps_vs_tau}), thereby making
the temperature dependence implicit. This representation of
the data allows us to illustrate more clearly several system-specific
aspects of the relationship between the structure and dynamics. As
expected, we find that the increase of $\tau$ correlates to that of
$N_\lps$. However, the spatial extension of LPS domains \textit{at
  fixed relaxation times} increases systematically with the fragility
$K$ of the model, which is estimated from fitting the relaxation times to
Eq.~\ref{eq:vft}. In the case of the more fragile WAHN mixtures, the
increase of relaxation times is evidently dictated by the growth of
LPS domains. Over the same range of $\tau$, the less fragile KA
mixtures show a weaker structural change, indicating that other
effects, such as dynamic facilitation, may also be playing an
important role. The overall trend of variation is consistent with the
correlation between the fragility and thermal rate of growth of LPSs put
forth in Ref.~\cite{coslovich_understanding_2007-1} and suggests that
the impact of static correlations on the dynamics should be more
pronounced the more fragile the liquid. We also note that the inclusion
of attractions tends to increase $N_\lps$ at fixed relaxation
times. This stabilization effect is in qualitative agreement with
recent observations on LJ and WCA fluids close to the triple
point~\cite{taffs_effect_2010}.


\section{Conclusions}

We have performed a crucial test on the dynamic role of static
correlations in glass-forming liquids by comparing two well-studied LJ
mixtures and their corresponding purely repulsive variants. Truncation
of the attractive part of the potential considerably shifts the glass
transition to lower temperatures and reduces the fragility of the
liquid, but does not alter the pair structure
significantly~\cite{berthier_nonperturbative_2009}. These phenomena
have been explained by resorting to indicators revealing more complex
structural correlations. Building on prior
work~\cite{coslovich_understanding_2007-1}, we have identified
correlated domains formed by locally preferred structures. We have
found that the growth, by decreasing the temperature, of LPS domains is
tightly connected to the onset of the slow dynamics regime. Furthermore,
an analysis of LPS domains has allowed us to clearly distinguish the
different dynamic behaviors of the LJ and WCAS models in terms of their
structure. In retrospect, these results suggest that even small
differences that are discernible at the level of pair correlations, may reflect
substantially different stages of the structural evolution of a
supercooled liquid and be associated with very different dynamic
regimes. A solution of the MCT equations for the dynamic correlation
functions---using two-body static correlations as input---only
partially accounts for the different dynamic behaviors of the LJ and WCA
models~\cite{voigtmann_2008,berthier_critical_2010}. Thus, proper
inclusion of many-body static correlations in theories of the glass
transition based on structural information seems crucial for a correct
description of the dynamics in fragile glass-formers. This suggests
revisiting and extending previous
attempts~\cite{kob_quantitative_2002} along these lines based on mode-coupling theory. Investigations of high order static correlations
extracted from simulations under amorphous boundary
conditions~\cite{biroli_thermodynamic_2008}, using patch
repetition~\cite{kurchan_levine_2011} or order mining
methods~\cite{Fang_Wang_Yao_Ding_Ho_2010}, as well as implementation
of alternative methods for LPS
determination~\cite{mossa_operational_2006,patrick_royall_direct_2008},
may provide further clues to improve our theoretical understanding of
the glass transition.

\begin{acknowledgements} 
  Useful discussions with L. Berthier, W. Kob, and G. Pastore are
  acknowledged.  D.C. acknowledge partial financial support by the
  Austrian Science Found (FWF) under Proj. No. P19890-N16.
\end{acknowledgements}


%

\end{document}